# ON THE POSSIBILITY OF A CONNECTION BETWEEN THE CONSTRUCTION OF A CLASS OF BIGEODETIC BLOCKS AND THE EXISTENCE PROBLEM FOR BIPLANES

**Frasser C.E.**
*Ph.D. in Engineering Sciences, Odessa National Polytechnic University, Ukraine*

**Abstract**
Graph theory and enumerative combinatorics are two branches of mathematical sciences that have developed astonishingly over the past one hundred years. It is especially important to point out that graph theory employs combinatorial techniques to solve key problems of characterization, construction, enumeration and classification of an enormous set of different families of graphs. This paper describes the construction of two classes of bigeodetic blocks using balanced incomplete block designs (BIBDs). On the other hand, even though graph theory and combinatorics have a close relationship, the opposite problem, that is, considering certain graph constructions when solving problems of combinatorics is not common, but possible. The construction of the second class of bigeodetic blocks described in this paper represents an example of how graph theory could somehow give a clue to the description of a problem of existence in combinatorics. We refer to the problem of existence for biplanes. A connection between the mentioned construction, the Bruck-Ryser-Chowla theorem and the problem of existence for biplanes is considered.

**Introduction**
   A *balanced incomplete block design* (or simply a block design) on a set $S$ with $|S| = n$, where $|S|$ is the cardinal number of $S$, is a family of subsets $B_1, B_2, \ldots, B_b$ of $S$ called *blocks* such that:
(a)  $|B_i| = k$, $1 \leq i \leq b$.
(b)  If $x \in S$, then $x$ belongs to exactly $r$ blocks $B_i$.
(c)  If $x, y$ are distinct elements of $S$, then $\{x, y\}$ is contained in exactly $\lambda$ blocks.
This block design is denoted $(b, n, r, k, \lambda)$.
  Just like any other combinatorial structure, block designs are defined in terms of certain parameters whose values determine the answer to the question of existence, that is, which values of these parameters produce the configuration in question and which do not. Given a $(b, n, r, k, \lambda)$-design, there are necessary conditions that its parameters must satisfy, namely, $bk = nr$, $r(k - 1) = \lambda(n - 1)$.
  The conditions previously described on the parameters of a block design are necessary, but not sufficient. It means that we can use them to rule out the existence of a block design for certain groups of parameters. However, being given the values of the parameters, which satisfy the conditions previously mentioned, does not guarantee the existence of a block design with those parameters. There are many groups of possible parameters for which the existence problem has not been settled.
  A design with $b = n$ is called *symmetric*. In such a design $r = k$ and hence such structure is called $(n, k, \lambda)$-design. For symmetric designs, there is an additional restriction for their existence [4, Theorem 3.1].

**Theorem 1.** (*Bruck-Ryser-Chowla*) Let $n$, $k$, $\lambda$ be integers for which there exists a symmetric $(n, k, \lambda)$-design. If $n$ is even, then $k - \lambda$ equals a perfect square. If $n$ is odd, then the equation

$$x^2 = (k - \lambda)y^2 + (-1)^{(n-1)/2}\lambda z^2$$

has a solution in integers $x$, $y$, $z$ not all zero. •

The set of vertices and edges of a graph $G$ are denoted $V(G)$ and $E(G)$, respectively. An *undirected graph* is one having edges with no direction. Two vertices of a graph $G$ are *adjacent* if they are connected by an edge. A graph in which every pair of its vertices is adjacent is called *complete*. The complete graph on $n$ vertices is denoted $K_n$. A *path* in a graph $G$ from vertex $v_0$ to vertex $v_n$ is a sequence $v_0v_1\ldots v_n$ of different vertices and is denoted $P(v_0, v_n)$. The number of edges of a path $P$ determines the length of this path and is represented by $|P|$. The length of a shortest path connecting vertices $u$ and $v$ in $G$ represents the *distance* between these two vertices. The greatest distance between any pair of vertices of $G$ is called the *diameter* of $G$, which is denoted $d(G)$. The number of edges incident to $v$ is called the *degree* of vertex $v$ and is denoted $deg(v)$. $G$ is said to be *regular of degree $k$* (or *$k$-regular*) if every vertex of $G$ has equal degree $k$ and *biregular* with *degree sequence* $(k, l)$ if for any vertex $v$ of $G$ $deg(v) = k$ or $deg(v) = l$ for fixed values $k$ and $l$, $k \neq l$. A *subgraph* $H$ of a graph $G$ is a graph whose vertices and edges are subsets of those of $G$. A *loop* is an edge that connects a vertex to itself. *Multiple edges* are two or more edges that are incident to the same two vertices.

In this research a graph is undirected, without loops or multiple edges. *Bigeodetic graphs* were defined by Srinivasan [5, p.102] as graphs in which each pair of nonadjacent vertices has at most two paths of minimum length between them. *K-geodetic graphs* have been defined in [1, p. 188] as graphs in which each pair of nonadjacent vertices has at most $k$ paths of minimum length between them. Thus, a $k$-geodetic graph is *geodetic* when $k = 1$, *bigeodetic* when $k = 2$, *trigeodetic* when $k = 3$, and so on. The minimum number of vertices whose deletion (implies also the deletion of the edges incident to the deleted vertices) disconnects $G$ is called *vertex connectivity* of a graph $G$. A graph $G$ is called *p-connected* if its vertex connectivity is equal to $p$. A *block* is a graph whose vertex connectivity $p$ is greater than 1. In [1, pp. 190-201], a general study of $k$-geodetic graphs has been performed and bigeodetic blocks have been considered there as a particular case of $k$-geodetic graphs.

A *cover* of a graph $G$ is a set $\{G_1, G_2, ..., G_m\}$ of complete subgraphs of $G$ such that $G_1 \cup G_2 \cup ... \cup G_m = G$. A cover of $G$ is called a $\Theta$-*cover* if any two elements of the cover are edge-disjoint.

Let $G$ be a graph having vertices $v_1$, $v_2$, ..., $v_n$. Let $A = \{G_1, G_2, ..., G_m\}$ be a cover of $G$, where $V(G_i) = \{v_{i_1}, v_{i_2}, ..., v_{j_i}\}$, $1 \leq i \leq m$. For each $i$, $1 \leq i \leq m$, take new vertices $v_{i_1 i}, v_{i_2 i}, ..., v_{i_{j_i} i}$ and construct a complete graph $K(G_i)$ on these vertices. Take $n$ new vertices $v_{10}$, $v_{20}$, ..., $v_{n0}$ and connect $v_{i_\ell i}$ to $v_{i_\ell 0}$ for $1 \leq \ell \leq j_i$, $1 \leq i \leq m$. The resulting graph is denoted $G^*(A)$.

Consider a $(b, n, r, k, \lambda)$-design on a set $S = \{x_1, x_2, ..., x_n\}$. Let $K_n$ be a complete graph with vertex set $\{x_1, x_2, ..., x_n\}$ and $G_i$ be a complete graph on vertex set of $B_i$, $1 \leq i \leq b$. Clearly, $A = \{G_1, G_2, ..., G_b\}$ is a cover of $K_n$. Construct graph $K_n^*(A)$ and denote it $K_n^*(r, k, \lambda)$. This is a $k$-connected, biregular block with degree sequence $(r, k)$. It has $n(r + 1)$ vertices and $nr(k + 1)/2$ edges. In Figure 2, $K_7^*(6,3,2)$ is constructed using the blocks of a $(14, 7, 6, 3, 2)$-design.

The described procedure to generate graph $K_n^*(r, k, \lambda)$ and the following theorem with its respective corollary are taken from [5, pp. 103-107].

**Theorem 2.** Let $\mu = \max[\max(|B_i \cap B_j| : i, j = 1,...,b, i \neq j), \lambda]$. Any pair of nonadjacent vertices of $K_n^*(r, k, \lambda)$ has at most $\mu$ distinct paths of minimum length between them. The diameter of $K_n^*(r, k, \lambda)$ is 4 if $B_i \cap B_j \neq \emptyset$ for every $i, j, i \neq j$ or 5 if $B_i \cap B_j = \emptyset$ for some pair of distinct values $i, j$. •

**Corollary 1.** If $(b, n, r, k, \lambda)$ is a symmetric design, then in $K_n^*(r, k, \lambda)$ there are at most $\lambda$ paths of minimum length between each pair of vertices. •

**Results**

Next, we present two constructions of bigeodetic blocks using block designs (Theorem 3 and Claim 1).

The construction described in Claim 1 has a special connotation because even though it describes a simple observation about the existence behavior pattern of the employed symmetric $((n^2+n+2)/2, n+1, 2)$-designs in a very short interval of integer values, the last section of this paper suggests that this simple observation could not be just a coincidence and could give a clue to the description of a more general problem of existence.

**Theorem 3.** For every $n \equiv 0$ or $1 \pmod{3}$, $n \geq 4$, there exists a bigeodetic block on $n^2$ vertices with diameter 4 or 5, with vertex connectivity 3 and degree sequence $(n-1, 3)$.

*Proof.* When $n \equiv 0$ or $1 \pmod{3}$, $n \geq 4$, there exists an $(n(n-1)/3, n, n-1, 3, 2)$-design on a set $S$ [2, Theorem 15.4.5]. Thus, taking $G_i$ to be a complete graph on vertices of $B_i$, $1 \leq i \leq n(n-1)/3$, graphs $G_1, ..., G_{n(n-1)/3}$ form a cover of the complete graph $K_n$ on vertex set $S$. Construct graph $K_n^*(n-1, 3, 2)$. This graph has $n^2$ vertices. According to Theorem 2, this is a bigeodetic graph of diameter 4 if $B_i \cap B_j \neq \emptyset$ for every $i, j, i \neq j$ or 5 if $B_i \cap B_j = \emptyset$ for some pair of distinct values $i, j$. It is easy to observe that $K_n^*(n-1, 3, 2)$ has degree sequence $(n-1, 3)$ and is 3-connected for $n \geq 4$. •

Next, we give the blocks of $(10, 6, 5, 3, 2)$ and $(14, 7, 6, 3, 2)$ designs, which are used to construct the bigeodetic blocks shown in Figures 1 and 2.

(i) $\{x_1, x_2, x_4\}$, $\{x_1, x_2, x_3\}$, $\{x_3, x_4, x_5\}$, $\{x_2, x_4, x_5\}$, $\{x_2, x_5, x_6\}$, $\{x_1, x_5, x_6\}$, $\{x_2, x_3, x_6\}$, $\{x_1, x_3, x_5\}$, $\{x_1, x_4, x_6\}$, $\{x_3, x_4, x_6\}$.

(ii) $\{x_1, x_2, x_4\}$, $\{x_1, x_2, x_3\}$, $\{x_3, x_4, x_6\}$, $\{x_3, x_4, x_5\}$, $\{x_2, x_5, x_6\}$, $\{x_3, x_6, x_7\}$, $\{x_1, x_6, x_7\}$, $\{x_1, x_4, x_7\}$, $\{x_2, x_3, x_7\}$, $\{x_1, x_3, x_5\}$, $\{x_2, x_5, x_7\}$, $\{x_2, x_4, x_6\}$, $\{x_1, x_5, x_6\}$, $\{x_4, x_5, x_7\}$.

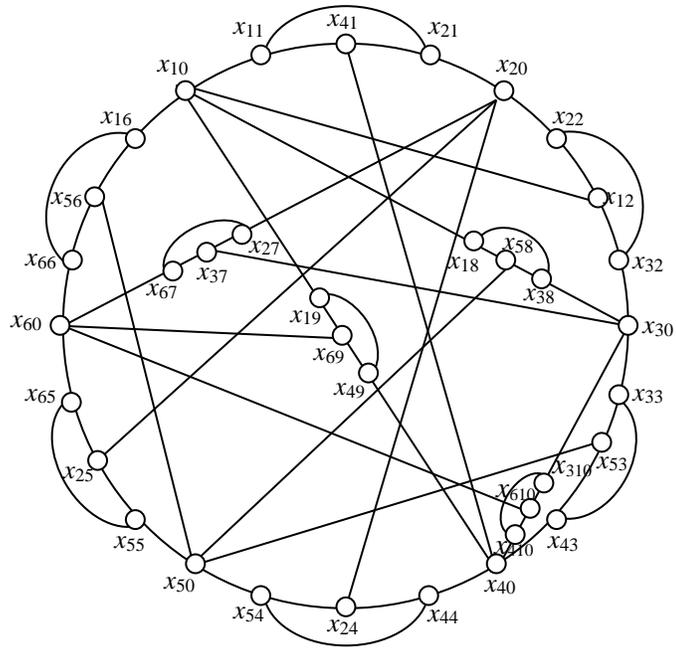

Fig. 1. A Bigeodetic Block generated by a (10, 6, 5, 3, 2)-design.

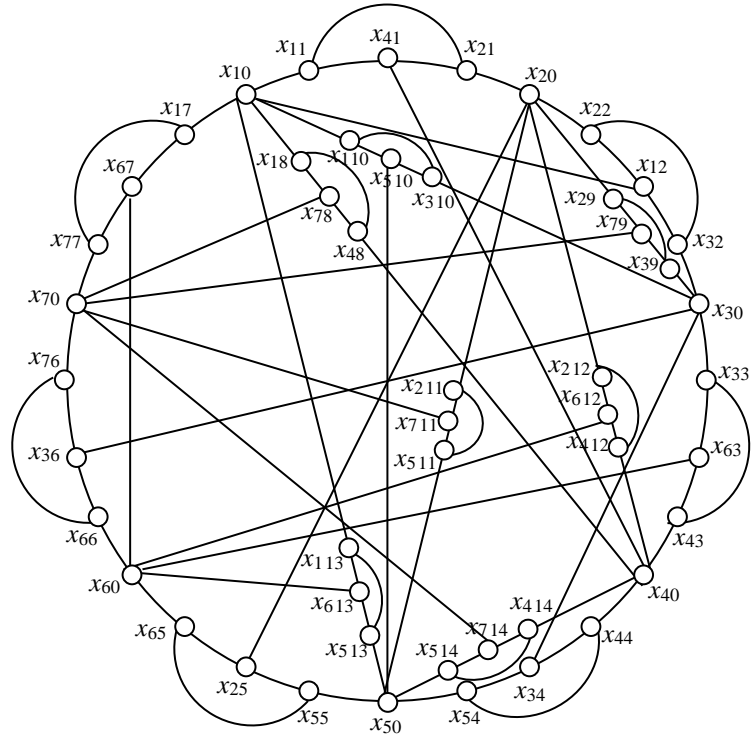

Fig. 2. A Bigeodetic Block generated by a (14, 7, 6, 3, 2)-design.

**Claim 1.** For every $n \equiv 1$ or $2 \pmod 4$, $2 \le n \le 10$, such that $(n-1)$ is a perfect square or $n \equiv 0$ or $3 \pmod 4$, $3 \le n \le 12$, such that $(n-1)$ is a prime power, there exists an $(n+1)$-regular, $(n+1)$-connected bigeodetic block of diameter 4. •

**Remark.** When $n \equiv 1$ or $2 \pmod 4$, $2 \le n \le 10$, such that $(n-1)$ is a perfect square or $n \equiv 0$ or $3 \pmod 4$, $3 \le n \le 12$, such that $(n-1)$ is a prime power, there exists a symmetric block design $((n^2+n+2)/2, n+1, 2)$ on a set $S$ with blocks $B_i$, $1 \le i \le (n^2+n+2)/2$ (the final section of this paper lists all symmetric $((n^2+n+2)/2, n+1, 2)$-designs so far found. Note that they obey the "simple pattern" of existence mentioned in the initial part of this remark). Let $G$ be a complete graph on vertex set $S$, and $G_i$ be a complete graph on vertex set $B_i$, $1 \le i \le (n^2+n+2)/2$. $G_1, ..., G_{(n^2+n+2)/2}$ form a cover of $G$. Construct graph $G^*_{(n^2+n+2)/2}(n+1, n+1, 2)$. This graph is an $(n+1)$-regular, $(n+1)$-connected one and has $(n^2+n+2)(n+2)/2$ vertices. According to Corollary 1, this is a bigeodetic block. Since any two blocks of a design $((n^2+n+2)/2, n+1, 2)$ have two common elements, the diameter of $G^*_{(n^2+n+2)/2}(n+1, n+1, 2)$ is 4. •

Next, we give the blocks of (4, 4, 3, 3, 2) and (7, 7, 4, 4, 2) designs, which are used to construct the bigeodetic blocks shown in Figures 3 and 4.

(i) $\{x_1, x_2, x_3\}$, $\{x_1, x_2, x_4\}$, $\{x_1, x_3, x_4\}$, $\{x_2, x_3, x_4\}$.

(ii) $\{x_1, x_2, x_3, x_4\}$, $\{x_1, x_3, x_5, x_7\}$, $\{x_1, x_4, x_5, x_6\}$, $\{x_1, x_2, x_6, x_7\}$, $\{x_2, x_3, x_5, x_6\}$, $\{x_2, x_4, x_5, x_7\}$, $\{x_3, x_4, x_6, x_7\}$.

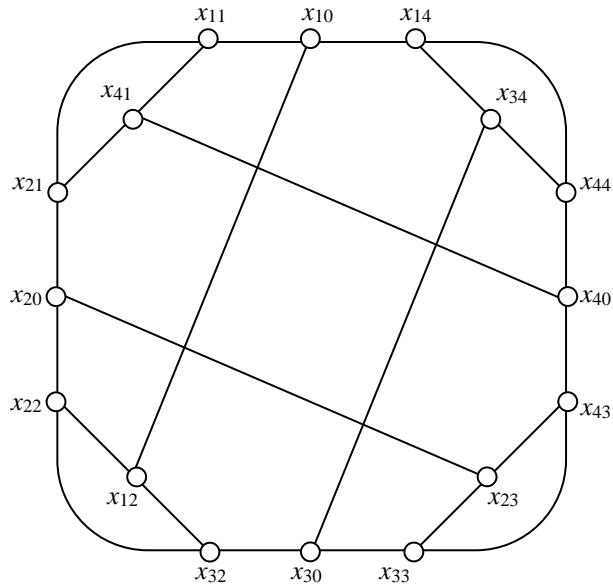

Fig. 3. A Bigeodetic Block generated by a (4, 4, 3, 3, 2)-design

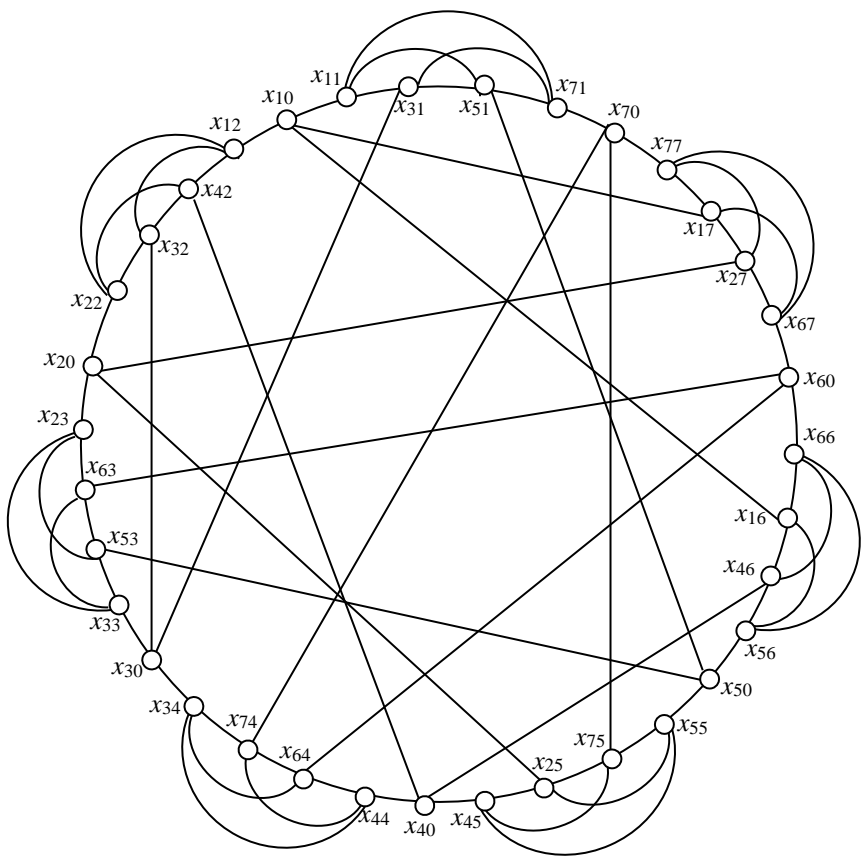

Fig. 4. A Bigeodetic Block generated by a (7, 7, 4, 4, 2)-design.

**Corollary 2.** If $(b, n, r, k, 2)$ is a block design, then $K_n^*(r, k, 2)$ is a bigeodetic graph of diameter either 4 or 5. •

Srinivasan, Opatrny, and Alagar [5, p. 111] considered that it is possible to construct $n$-regular, $k$-connected bigeodetic blocks of diameter $d$, where $k, n, d \geq 2$, $k \leq n$.

They have denoted this class of blocks $B(k, n, d)$ and have posed the following problem:

Is class $B(k, n, d)$ nonempty for every $k, n, d \geq 2$, $k \leq n$?

**Claim 2.** For every $n \equiv 1$ or $2 \pmod 4$, $2 \leq n \leq 10$, such that $(n-1)$ is a perfect square or $n \equiv 0$ or $3 \pmod 4$, $3 \leq n \leq 12$, such that $(n-1)$ is a prime power, class $B(n+1, n+1, 4)$ is nonempty. •

**Concluding Remarks**

Similar constructions to those ones described in Theorem 3 and Claim 1 for bigeodetic blocks can be formulated for trigeodetic blocks using $(b, n, r, k, 3)$-designs. Thus, when $n \equiv 1 \pmod 2$, $n \geq 5$, there exists an $(n(n-1)/2, n, 3(n-1)/2, 3, 3)$-design [2, Theorem 15.4.5]. In the same way, when $n \equiv 0$ or $2 \pmod 3$, $3 \leq n \leq 14$, $n \neq 12$, there exists a symmetric $((n^2+n+3)/3, n+1, 3)$-design, namely: (5, 4, 3), (11, 6, 3), (15, 7, 3), (25, 9, 3), (31, 10, 3) [2, Appendix 1], (45, 12, 3), (71, 15, 3) [3, p. 105], which is called a *triplane*.

For any fixed integer value $\lambda \geq 2$, the question of whether there exists an infinite number of symmetric $(n, k, \lambda)$-designs is unresolved. In particular, when $\lambda = 2$, such a design is called a *biplane* and there exists only a few known examples, namely, (4, 3, 2), (7, 4, 2), (11, 5, 2), (16, 6, 2), (37, 9, 2), (56, 11, 2), (79, 13, 2). The first two biplanes are here used to generate two bigeodetic blocks (see Figures 3 and 4). Ryser [4, pp.114-115] proved that if in a symmetric $(n, k, \lambda)$-design $n$ is odd and $(k, \lambda) = 1$ where $(k, \lambda)$ denotes the positive greatest common divisor of $k$ and $\lambda$, then $(k - \lambda, \lambda) = 1$ and the equation $x^2 = (k - \lambda)y^2 + (-1)^{(n-1)/2}\lambda z^2$ associated with the Bruck-Ryser-Chowla theorem has a solution in integers $x$, $y$, and $z$, not all zero. It is evident that for $n \equiv 0$ or 3 (mod 4) with $n > 3$ and $(n-1)$ a prime power, $(n^2+n+2)/2$ is odd and $((n^2+n+2)/2, n+1, 2)$-biplanes satisfy the conditions established by Ryser. As a result, when substituting $n$, $k$, and $\lambda$ into $x^2 = (k - \lambda)y^2 + (-1)^{(n-1)/2}\lambda z^2$ for $(n^2+n+2)/2$, $n+1$, and 2, respectively, an equation with a solution in integers $x$, $y$, and $z$, not all zero is generated. Consequently, for $n \equiv 0$ or 3 (mod 4) with $n > 3$ and $(n-1)$ a prime power, $(n^2+n+2)/2$ is odd and $((n^2+n+2)/2, n+1, 2)$-biplanes satisfy the necessary condition established in Theorem 1 for their existence.

Due to the fact that some special families of block designs turned out to be finite, it is believed that only finitely many symmetric designs exist for any fixed $\lambda > 1$. Assuming that this is true, one could formulate the question if for a given finite integer interval, biplanes and their existence respond to the same simple pattern of behavior described in the construction of the Claim 1 bigeodetic blocks.

*Assume that n belongs to a finite interval of integer values and m is a fixed integer, $m \geq 12$. Could it be possible that being $n \equiv 1$ or 2 (mod 4), $2 \leq n < m$, such that $(n-1)$ is a perfect square or $n \equiv 0$ or 3 (mod 4), $3 \leq n \leq m$, such that $(n-1)$ is a prime power, there exists a symmetric block design $((n^2+n+2)/2, n+1, 2)$?*

Note that the answer to this question is in the affirmative for $m = 12$.